\documentclass[a4paper]{spie}  
\usepackage{amsmath,amsfonts,amssymb}

\usepackage[utf8]{inputenc}   
\usepackage[english]{babel}   
\usepackage[usenames,dvipsnames]{xcolor} 
\usepackage[colorlinks=true, allcolors=blue, pdftex]{hyperref}
\usepackage{aas_macros}
\usepackage[]{siunitx}
\usepackage{paralist}
\usepackage{graphicx}
\usepackage[frozencache,cachedir=.]{minted}

\title{Toward the remotization and robotization \\ of the OARPAF telescope}

  \author[a,*]{Davide~Ricci}
  \author[a]{Lorenzo~Cabona}
  \author[b,c]{Silvano~Tosi}
  \author[d]{Sandro~Zappatore}
  
  \affil[a]{INAF-Osservatorio Astronomico di Padova, Vicolo
    dell'Osservatorio 5, 35122 Padova, Italy.}

  \affil[b]{Università degli Studi di Genova, DIFI Dipartimento di
    Fisica, Via Dodecaneso 33, 16146, Genova, Italy.}

  \affil[c]{INFN-Sezione di Genova, Via Dodecaneso 33, 16146 Genova,
    Italy.}
  
  \affil[d]{Università di Genova, DITEN Dipartimento di Ingegneria
    delle Telecomunicazioni, Elettrica, Elettronica e Navale, Via
    all’Opera Pia 11A, 16145, Genova, Italy.}

  \authorinfo{$^*$Contact information: davide.ricci@inaf.it}

\begin{document}
\maketitle

\renewcommand{\thefootnote}{\alph{footnote}}

\begin{abstract}
  OARPAF (Osservatorio Astronomico Regionale Parco Antola Comune di
  Fascia, Italy) hosts an 80cm Astelco telescope with a Gambato Dome,
  SBIG-STX camera, Davis weather station, and SBIG AllSky camera. We
  present a layer-structured \texttt{python3} framework to control
  these devices.
  Layer 1 provides straightforward getter/setter interface for
  ``atomic'' operations on devices.
  Layer 2 wraps the above mentioned atomic operations into
  ``ESO-style'' Templates, to perform sequences of common pointing,
  observation, and calibration operations called ``Observation Blocks''
  (OBs) that are run by a sequencer.
  Layer 3 is a REST API based on HTTP verbs to expose methods that
  control Layer 1 devices and Layer 2. We also present a web interface
  built on top of this layer.
  The work is part of the frame for remoting and robotizing the
  observatory.
\end{abstract}




\section{Introduction}
\label{sec:intro}

The Regional Astronomical Observatory of the Antola Park (OARPAF),
located in the Ligurian Apennines near the city of Genova, Italy, is
operational since years not only for educational~\cite{Righi2015,
  Cabona2016, Nicolosi2019} and general public
activities\cite{2012ASInC...7....7F}, but also for scientific
observations, which led to results in the field of light curves of
quasars\cite{2016NCimC..39..284R} and exoplanetary transits \cite{
  2017PASP..129f4401R, 2022MNRAS.509.1447M}.

These activities increased the interest in the facility by local
institutions that contributed to its technical improvement
\cite{2021RMxAC..53...14R}, and paved the path for the future
development of a multi-purpose instrument capable of providing three
focal stations at the scientific Nasmyth focus of the telescope: an
imager, a long slit spectrograph, and an échelle
spectrograph\cite{2020SPIE11447E..5JC}.

The commissioning of the instrumentation and launch of the new phase
of the scientific exploitation of OARPAF, as well as the future
perspective, are described in a recent
publication\cite{2021JATIS...7b5003R}.
However, the development and evolution of the facility followed a
non-linear path, which ultimately led to a wide range of hardware and
software interfaces that make the remotization and robotization of the
observatory a long but challenging task, starting from instrument
control software development.

In Sect.~\ref{sec:devs} we present the hardware devices to be remotely
controlled.  In Sect.~\ref{sec:layers} we present a layer-structured
\texttt{python3} framework to control the observatory devices.
Conclusions are drawn in Sect.~\ref{sec:conc}.

\label{fig:devs}

\section{Devices}
\label{sec:devs}

\label{fig:panel}

The observatory comprises several hardware devices. Some of the
devices share the same control electronics and the same software
interface for their operation, so they can be treated as a
subsystem. In the following description we summarize these subsystems
and individual devices.

\begin{description}
  
\item[Dome subsystem] To overcome the problems related to the original
  dome of the observatory, a new dome was installed in 2020 by the
  Gambato company\footnote{\url{https://www.gambato.it/}}.
  Its control electronics consists of a proprietary module called
  Observatory Control System (OCSIII) by the OmegaLab
  company\footnote{\url{http://atcr.altervista.org/ita/index.html}},
  based on \texttt{sparkfun} \texttt{STM32} board.  The module also
  includes a set of relays for controlling a dome light and a flat
  field lamp to illuminate a white screen on the dome.
  The OCSIII is interfaced by means of a commercial control software,
  named \textit{Ricerca}, developed by OmegaLab, which operates under
  Windows and provides users a set of Graphical User Interfaces
  (GUIs). The communication protocol is proprietary and
  undocumented. However, \emph{Ricerca} makes use of Astronomy Common
  Object Model (ASCOM) drivers, an open initiative to provide a
  standard interface to a wide range of astronomy equipment including
  mounts, focusers, and cameras in a Microsoft Windows \texttt{.NET}
  environment.
  These components are shown as \textsf{Dome}, \textsf{Lamp} and
  \textsf{Light} devices in Fig.~\ref{fig:devs}, while the
  corresponding ASCOM drivers are indicated as \texttt{OCS Dome} and
  \texttt{OCS Switch} in the same figure.

\item[Telescope subsystem]
  The OARPAF telescope is a $0.8$m 
  alt-azimuthal Cassegrain-Nasmyth T0800-01 telescope manufactured by
  ASTELCO Systems\footnote{\url{http://www.astelco.com}}.
  Its control electronics consists of a PLC-based system which can be
  operated with getter/setter commands sent via \texttt{telnet} using
  the proprietary \texttt{OpenTSI} protocol.
  The interface for this device is the \textsc{AsTelOS} proprietary
  software provided by Astelco Systems, with GUIs for Windows and
  Linux operating systems.
  Astelco also provides ASCOM drivers which allow controlling the
  telescope, the motorized secondary mirror for focus adjustments, and
  the field de-rotator on one of the Nasmyth foci, which is used for
  scientific operations.
  These components are denoted as \textsf{Telescope}, \textsf{Rotator}
  and \textsf{Focuser} devices in Fig.~\ref{fig:devs} and the
  corresponding ASCOM driver is indicated as \texttt{ttTCLM multi} in
  the same figure.

\item[Camera subsystem] Currently, the scientific detector is a
  \textsc{sbig stx 16801} camera with a Class-1 CCD. This camera is
  equipped with a Kodak \textsc{kaf-16801} sensor having 16 million
  $9\mu$m pixels, a \textsc{fw7-stx} filter wheel with standard,
  $50$mm $UBVRI$ and $H\alpha$ filters, a \textsc{stx}-Guider, and the
  new \textsc{ao-x} module for tip-tilt correction, all provided by
  the Diffraction Limited
  company\footnote{\url{https://diffractionlimited.com/}}.
  The electronics controlling these devices is integrated in the
  camera component and provides both Universal Serial Bus (USB) and
  Ethernet at the physical layer. We use the latter in our setup.
  The Ethernet interface allows us to access a basic set of
  Hypertext Transfer Protocol (HTTP) Common Gateway Interface (CGI)
  scripts for setting and getting camera parameters and for launching
  image acquisitions.
  These components are shown as the \textsf{Camera} device in
  Fig.~\ref{fig:devs}.

\item[Meteo station] Weather monitoring is carried out by means of a
  Davis Vantage Pro 2 meteo station manufactured by the Davis
  Instruments
  company\footnote{\url{https://www.davisinstruments.com/}}.  The
  station is installed on a pole close to the dome. A radio link
  transfers data to a control panel, placed under the dome, which
  archives them on a small memory inside to the panel itself.
  An additional Davis component, called Data Logger, provides an
  Ethernet interface, employed to access the control pannel via a TCP
  connection.
  This component is shown as the \textsf{Meteo} device in
  Fig.~\ref{fig:devs}.

\item[Allsky Camera] A SBIG Allsky Camera is co-located with the meteo
  station in order to monitor the sky coverage. The camera is
  commercialized by Diffraction Limited group.
  The device adopts a documented protocol and transmits data over a
  V.24 (aka RS-232) serial interface. A Moxa NPort 5110 server permits
  creating a virtual serial interface on the computer devoted to grab
  images from the camera.
  The camera is shown as the \textsf{AllSky} device in
  Fig.~\ref{fig:devs}.
   
\item[IP camera] The dome and telescope can be visually monitored with
  \textsc{D-Link dcs-5222l} IP camera that allows us to control the
  Pan and Tilt via the Open Network Video Interface Forum (ONVIF)
  protocol. The device is Ethernet-connected to the internal network.
  This component is interfaced via HTTP CGI scripts. 
  This component is shown as the \textsf{Ipcam} device in
  Fig.~\ref{fig:devs}.
  
\item[Smart switches] Several smart switches are responsible to
  control the power status of specific subsystems. For instance, the
  camera subsystem and the control cabinet subsystem are switched on
  or off by these devices, named \textsf{Sonoff} and \textsf{PLC
    Cabinet + modebus} in Fig.~\ref{fig:devs}.
  
\end{description}

These subsystems and single devices are in principle independent. A
first, basic goal for the automatization of an observatory consists of
\emph{minioning} the dome to the telescope, in order for the dome to
follow the telescope pointing and the subsequent tracking.
Since both the dome and the telescope expose their own ASCOM drivers,
we decided to accomplish this task by means of a client component,
called \textsf{ASCOM Device Hub}, which also re-exposes the drivers.
(see Fig.~\ref{fig:devs}).
 
ASCOM is a \emph{de facto} standard (proposed early 2000) for a
significant number of Commercial Off-the-Shelf (COTS) components and
it widely used for small private observatories. However, it is neither
cross-platform nor remotely accessible. For these reasons, a parallel,
compatible protocol called \texttt{ASCOM Alpaca} was recently
introduced. Alpaca consists of a Representational State Transfer
(REST) HTTP Application Programming Interface (API) to enable ASCOM
applications and devices to communicate across modern network
environments. Alpaca allows HTTP \texttt{GET} and \texttt{PUT}
requests, as getter/setter calls, to be forwarded to endpoints that
mimic the corresponding ASCOM class methods.

Up to date, native Alpaca drivers are still uncommon.
In order to use this paradigm to expose traditional \texttt{ASCOM}
drivers, a web server component called \textsf{ASCOM Remote} was
introduced.  We decided to expose all the \texttt{ASCOM}-based devices
of OARPAF through this component (Fig.~\ref{fig:devs}).
In this way, everything is accessible via the Local Area Network (LAN)
of the observatory: this represents the baseline of the layered
\texttt{python3} control software described in the following section.

\section{Software layers}
\label{sec:layers}

\begin{figure} [t]
  \centering
  \includegraphics[width=\textwidth]{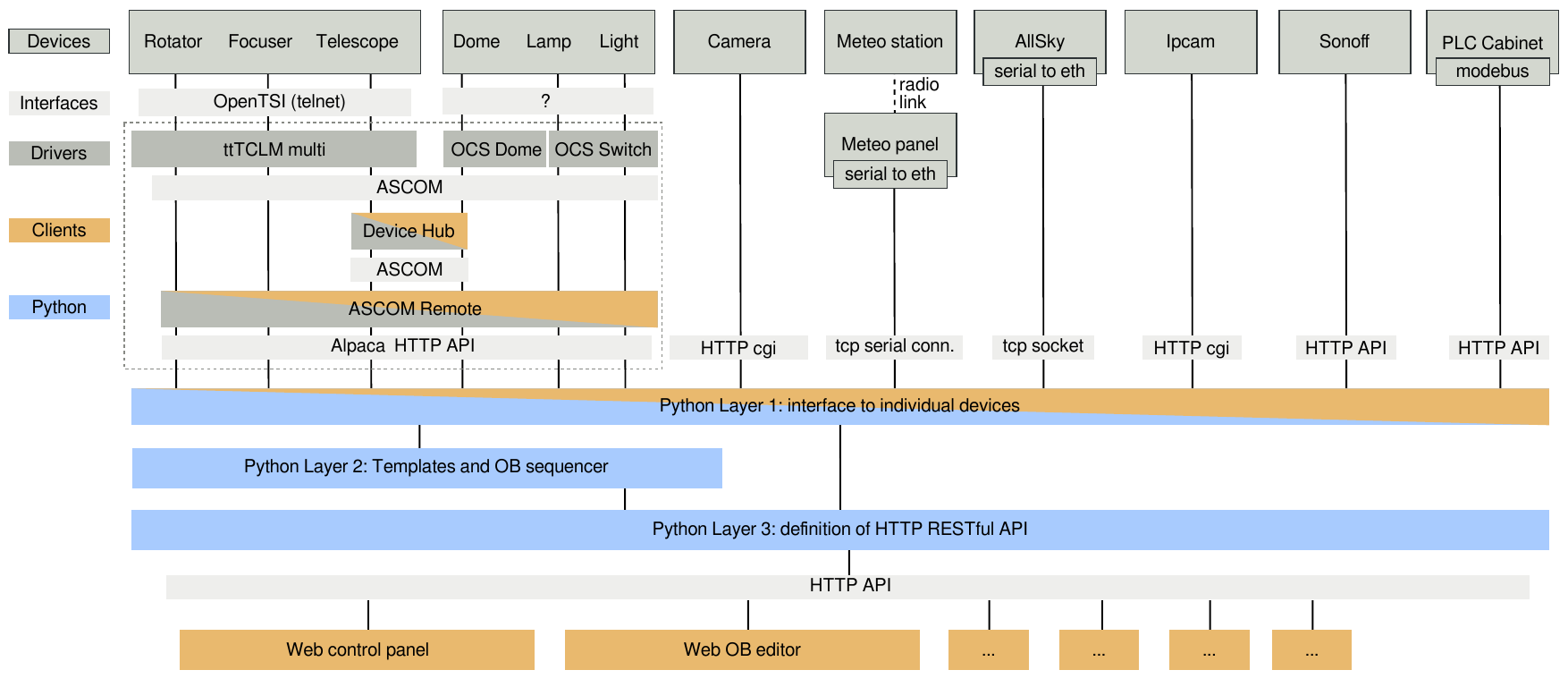}
  \caption{ Schema of OARPAF components showing ``subsystem'' devices,
    their software intefaces, the drivers, and the python layered
    software described in Sect.~\ref{sec:l3}.The dashed line shows the
    ASCOM part, that needs to be managed by a buffer Windows pc.  }
\end{figure}

We started building a layered \texttt{python3} instrument control
software that manages all these LAN-accessible resources.

\subsection{Layer 1}
\label{sec:l1}

Layer 1 provides a straightforward interface for ``atomic'' operations
on individual devices in an easy way: a class and eventually inherited
classes were implemented for each subsystem by exploiting the python
\texttt{@property} decorator to permit getter/setter to access the
command class attributes.
For example, to get the dome position and to set a new one, the
following HTTP Alpaca calls:
\begin{minted}{bash}
curl -X GET "http://server/api/v1/dome/0/azimuth" -H  "accept: application/json" # returns 123

curl -X PUT "http://server/api/v1/dome/0/slewtoazimuth" -H "accept: application/json" 
     -H  "Content-Type: application/x-www-form-urlencoded" -d "Azimuth=234"
\end{minted}
are implemented into python methods of the \texttt{Dome} inherited
class for our Ascom devices:
\begin{minted}{python}
class Dome(AscomDevice):
...
    @property
    def azimuth(self):
        res = self.get("azimuth")
        self._azimuth = res
        return self._azimuth

    @azimuth.setter
    def azimuth(self, a):
        data = {"azimuth": a}
        res = self.put("slewtoazimuth", data=data)
        self._azimuth = res
\end{minted}
where \texttt{self.get()} and \texttt{self.put()} manage the
corresponding calls, handle and log errors.
Then, the azimuth can be accessed and set in python with
\begin{minted}{python}
dome = Dome()
az = dome.azimuth   # get the value 123
dome.azimuth = 234  # set a new value
\end{minted}
This pattern is kept for all the OARPAF subsystems.
Layer 1 allows a straightforward command-line control of the
instrument using interactive shells such as \texttt{ipython} and, of
course, it can be easily implemented in scripts.

\subsection{Layer 2}
\label{sec:l2}

Layer 2 wraps the atomic operations of Layer 1 into scripts to carry
out the most common operations, such as, for example, taking
calibration frames, pointing the telescope, launching an observation.

It is the Template approach, widely used in large observatories such
as the European Southern Observatory (ESO), and recently adopted for
the instrument software development of the forthcoming SHARK-NIR
instrument\cite{shark-ricci} for the Large Binocular Telescope (LBT),
presented in this conference.

These scripts normally involve more than one subsystem at a time. For
example, a sequence of bias frames requires not only a camera set up,
but also dome lights off, and eventually the telescope petals closed.

Template scripts are developed as python modules containing a single
inherited class and a single method.
The base template class, from which the other templates inherit,
contains methods responsible for parameter checking, errors handling,
and the graceful abort.

In the following example, a test template to switch off all light
sources is shown.
\begin{minted}{python}
from utils.logger import log
from devices import lamp, light
from templates.basetemplate import BaseTemplate

class Template(BaseTemplate):
    def __init__(self):
        super().__init__()
        self.name = "lampsoff"
        self.description = "Switches off all the lamps"

    def content(self, params={}):
        for source in [lamp, light]:
            while source.state != False:
                log.debug(f"{source.description} still on")
                source.state = False
            log.info(f"{source.description} is now off")
        return  
\end{minted}
In this example, the template takes no parameters. In general,
parameters are stored in a JavaScript Object Notation (JSON) file with
a default setup.
The observer can use these default JSON files as blueprints to build
sequences of operations called ``Observation Blocks'' (OBs). In this
example, we show an OB composed by two templates: the acquisition of
the stellar field of a Messier object, and its observation:
\begin{minted}{json}
[
    {
        "template": "acquisition",
        "params": {
            "radec": [ 13.21, 18.16 ],
            "offset": [ 0, 0 ]
        }
    },
    {
        "template": "observation",
        "params": {
            "objname": "M53",
            "binning": 1,
            "filter": "V",
            "exptime": 600,
            "repeat": 3
        }
    }
]  
\end{minted}
OBs are run by a sequencer script, that takes in input the JSON file,
validates it, and calls in sequence the templates, passing the
specified parameters. For example: \texttt{./sequencer.py m53.json}.

\subsection{Layer 3}
\label{sec:l3}

Layer 3 consists in a RESTful API based on HTTP verbs manage the
editing the Layer 2 OBs, but also to expose methods that manage the
Layer 1 devices.
The API pattern follows the best practices of this approach, for
instance:
\begin{inparaitem}
\item \texttt{GET}: Get a resource, or the state of an action.
\item \texttt{POST}: Create a new resource, or start an action.
\item \texttt{PUT}: Update a resource.
\item \texttt{DELETE}: Delete a resource, or stop an action.
\end{inparaitem}
It is realized by means of the \texttt{flask-restx} python package,
that allows Swagger\footnote{\url{https://swagger.io/}} auto-generated
documentation and test benchmarks directly from the python code.

This allows principally to create and modify OBs, the same way ESO P2
APIs\footnote{\url{https://www.eso.org/copdemo/apidoc/index.html}} do.
Moreover, it is possible to call the sequencer to launch OBs, and
check/manage the execution status.

Taking advantage from this approach, we also decided to create
endpoints for our individual methods. For Alpaca devices this means in
a certain way to go back by one step, but Alpaca APIs do not
completely follow RESTful patterns, they just map 1:1 ASCOM
methods. We provide instead a homogeneous approach for all OARPAF
subsystems. Moreover, other subsystems such as \textsf{Meteo} and
\textsf{AllSky} can be accessed the same way as others.

APIs allow in principle to build any kind of client
interface.
In Fig.~\ref{fig:panel} we show a draft of a web-based client
interface built on top of Layer 3 using the \texttt{flask} python
package and the \texttt{bootstrap5} suite, which allow to monitor and
control most of OARPAF subsystems.
Other possible clients include, for example, simplified touch
applications for children didactics, monitoring databases, or
specialized automatic schedulers for robotic operations.

Scheduling is an important task for robotic operations: the role of
the scheduler consists in drawing a mathematically optimal planning of
the OB observations in order to maximize the scientific return of the
observatory.
The scheduler firstly compute the observability of sets of targets
imposing different constraints (e.g airmass, Moon separation, Moon
illumination), then it designs the schedule for the night planning the
observation of the OB when it guarantees an high scientific
measurement quality, and minimizing the the dead time between
observation configurations (mostly telescope and dome slews).
To achieve this, we plan to develop the OARPAF scheduler following the
approach outlined in a recent publication \cite{cabona2021scheduling}.

\begin{figure} [t]
  \centering
  \includegraphics[width=\textwidth]{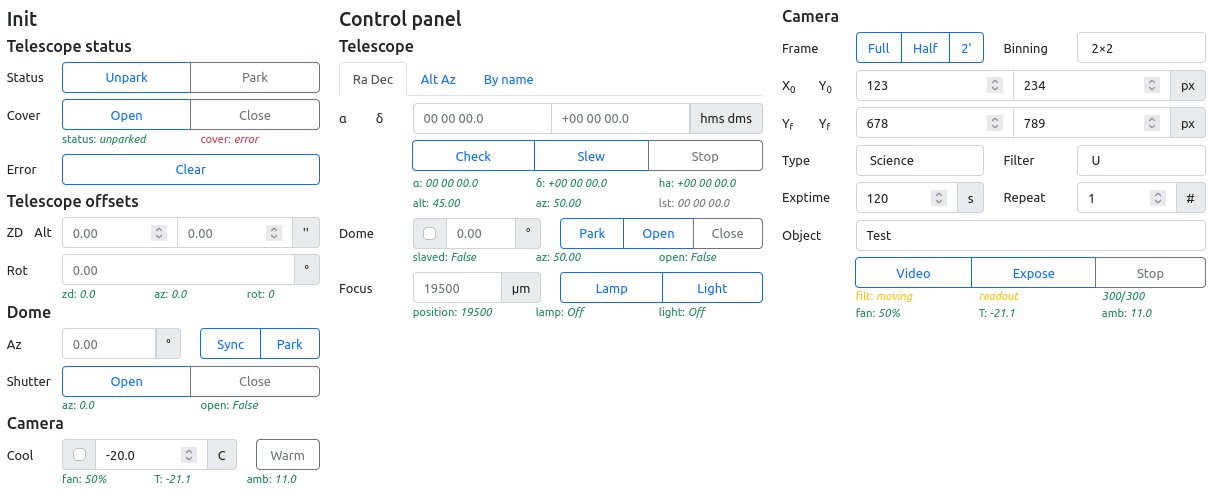}
  \caption{ Draft of OARPAF Control panel.  }
\end{figure}

\section{Conclusion}
\label{sec:conc}

In this work we presented a layered-structured instrument control
software for the OARPAF 0.8m-class observatory based on the python
programming language.
The main effort consisted in a first hardware setup to make all the
resources accessible to the LAN. Then, we provided to interface every
resource with a getter/setter approach for ``atomic'' operations on
each device.
These operations are implemented into Templates scripts, that are
called individually or in Observation Block sequences by passing a
JSON file containing the corresponding parameters.
A modern, self-documented RESTful HTTP API allows ``\emph{ESO
  P2-style}'' OB management, individual device control, and offers to
the community a tool for the realization of client interfaces.
We developed a first client interface based on web technologies
consisting in a general observatory control panel to simplify the
remote control of the observatory.  API will be used to enhance
automatic operation towards the complete robotization of the
structure.

\acknowledgments

We are deeply appreciative to Associazione Urania for general public
activities. We thank Regione Liguria, Comune di Fascia and Ente Parco
Antola for the always fruitful collaboration.  Fundings for the
facility and instruments were provided by Regione Liguria, Programma
Italia-Francia Marittimo, Comune di Fascia, Ente Parco Antola,
Università di Genova, DIFI and DIBRIS. Instruments for outreach events
and activities for students received contributions by Piano Lauree
Scientifiche (PLS) of MIUR and ``Departments of Excellence''.
Individuals have received support by INAF, INFN and MIUR (FFABR).

\bibliographystyle{spiebib} 
\bibliography{biblio} 

\end{document}